\documentclass[conference]{IEEEtran}
\IEEEoverridecommandlockouts
\usepackage{cite}
\usepackage{amsmath,amssymb,amsfonts}
\usepackage{algorithmic}
\usepackage{graphicx}
\usepackage{textcomp}
\usepackage{xcolor}
\usepackage{algorithm}
\usepackage{array}

\usepackage[caption=false,font=normalsize,labelfont=sf,textfont=sf]{subfig}
\usepackage{stfloats}
\usepackage{url}
\usepackage{verbatim}

\usepackage{subfloat}

\def\BibTeX{{\rm B\kern-.05em{\sc i\kern-.025em b}\kern-.08em
    T\kern-.1667em\lower.7ex\hbox{E}\kern-.125emX}}
\begin{document}

\title{Clustering-Based User Selection in Federated Learning: Metadata Exploitation for 3GPP Networks\\
\vspace{-2mm}
\thanks{ 
Chen Sun is the corresponding author: chen.sun@sony.com
}
}

\author{\IEEEauthorblockN{
Ce Zheng\IEEEauthorrefmark{1}, Shiyao Ma\IEEEauthorrefmark{2}, Ke Zhang\IEEEauthorrefmark{3}, Chen Sun\IEEEauthorrefmark{4}, Wenqi Zhang\IEEEauthorrefmark{4}
}

\IEEEauthorblockA{\IEEEauthorrefmark{1}Department of Broadband Communication, Pengcheng Laboratory, Shenzhen, China}
\IEEEauthorblockA{\IEEEauthorrefmark{2}College of Computer and Information Science, Southwest University, Chongqing, China}
\IEEEauthorblockA{\IEEEauthorrefmark{3}Intelligent Software Laboratory, Waseda University, Tokyo, Japan}
\IEEEauthorblockA{\IEEEauthorrefmark{4}Research and Development Center, SONY(China) Ltd, Beijing, China}
\IEEEauthorblockA{Email: zhengc@pcl.ac.cn, shiyaoma.cs@gmail.com, zhangke@moegi.waseda.jp,\\ chen.sun@sony.com, wenqi.zhang@sony.com}
}

\maketitle

\begin{abstract}
Federated learning (FL) enables collaborative model training without sharing raw user data, but conventional simulations often rely on unrealistic data partitioning and current user selection methods ignore data correlation among users. To address these challenges, this paper proposes a metadata-driven FL framework. We first introduce a novel data partition model based on a homogeneous Poisson point process (HPPP), capturing both heterogeneity in data quantity and natural overlap among user datasets. Building on this model, we develop a clustering-based user selection strategy that leverages metadata, such as user location, to reduce data correlation and enhance label diversity across training rounds. Extensive experiments on FMNIST and CIFAR-10 demonstrate that the proposed framework improves model performance, stability, and convergence in non-IID scenarios, while maintaining comparable performance under IID settings. Furthermore, the method shows pronounced advantages when the number of selected users per round is small. These findings highlight the framework’s potential for enhancing FL performance in realistic deployments and guiding future standardization.
\end{abstract}

\begin{IEEEkeywords}
federated learning, user selection, metadata, clustering, 3GPP.
\end{IEEEkeywords}

\section{Introduction}
Federated learning (FL) is a distributed machine learning paradigm that enables multiple users to collaboratively train a global model without sharing their raw data~\cite{mcmahan2017communication}. Each user performs local training on its own dataset, and a central server aggregates the local model updates to obtain the global model. This privacy-preserving framework has attracted increasing attention in data-sensitive applications such as Internet of Things (IoT), vehicular networks, and mobile edge computing.
However, existing studies overlook two key issues in FL:
(1) the commonly used data partition methods fail to
realistically simulate user data distribution in federated
scenarios, and (2) user selection strategies often neglect the
impact of metadata and the resulting data correlation among
users.

\textbf{Data Partition}: 
Unlike centralized learning, where the entire dataset is available on the server, FL operates over multiple users, each with its own dataset and distinct data distribution. To simulate such federated scenarios, many studies adopt artificial partitions of standard datasets, such as MNIST~\cite{lecun1998gradient} and CIFAR-10~\cite{krizhevsky2009learning}. McMahan~\cite{mcmahan2017communication} proposed two partitioning schemes: IID, where data are shuffled and equally distributed among clients, and non-IID, where each client holds only a few classes (e.g., two shards). However, these methods fail to capture quantity skew, i.e., variations in the number of samples across users. 
Yurochkin~\cite{yurochkin2019bayesian} used a Dirichlet distribution to simulate heterogeneous data distributions. While more flexible, this approach requires parameter tuning and assumes independent datasets across users, which may not reflect realistic scenarios with overlapping data. In~\cite{sun2022user}, data correlation was considered using the camera field of view (FoV) in V2X networks, where the overlap ratio between FoVs quantifies data correlation between devices. Nevertheless, this method requires prior calculation of correlations, which becomes impractical as the number of users increases, and it still neglects heterogeneity in data quantity.

\textbf{User Selection:} In large-scale FL systems, such as IoT networks, it is impractical for the server to communicate with all users due to limited spectrum and communication resources. Consequently, only a subset of users can participate in each training round. Moreover, users possess heterogeneous data quantities and qualities, making user selection a critical concern in FL.
Various user selection strategies have been proposed. Choong~\cite{cho2020client} showed that selecting clients with higher local losses can accelerate convergence and improve FL performance, but this approach introduces additional communication and computational overhead. Ma~\cite{ma2021client} grouped clients to make the label distribution within each group closer to the global label distribution, improving performance at the risk of privacy leakage. Wang et al.~\cite{wang2020optimizing} proposed a deep reinforcement learning–based framework to select the optimal subset of devices and mitigate the non-IID problem; however, this requires the central server or base station to communicate with all devices in every round, which is impractical for large-scale systems. Although these works focus on selection criteria such as local losses or label distributions, they overlook a more fundamental question: what causes the ``data problem'' among users? Our intuitive analysis suggests that users' raw data are inherently linked to their metadata.

To address the limitations discussed above, we first propose a new data partition model for realistic FL simulation. Inspired by the data collection mechanisms in sensor networks~\cite{kalor2018random}, user locations follow a uniform distribution, while data samples follow a homogeneous Poisson point process (HPPP). Each data point corresponds to one data sample, and a user captures the sample if it falls within its sensing radius. In this way, the model naturally introduces data correlation among users with overlapping coverage areas, while also reflecting heterogeneity in data quantity.
Building on the improved data partition model, we then propose a general user selection strategy that exploits metadata to mitigate data correlation in FL systems. Specifically, users are first clustered based on metadata, and one user is selected from each cluster. This strategy is based on the assumption that users with similar metadata are highly correlated or exhibit similar data distributions. To the best of our knowledge, only our previous work~\cite{chen2020joint} considered a minimum separation distance for user selection to mitigate the negative effects of data correlation in V2X networks. In addition, from a standardization perspective, user location information has been considered in 3GPP SA2 specifications for supporting FL applications~\cite{3gpp_tr22874,3gpp_tr23700_80, 3gpp_tr22876}. The proposed metadata-based user selection aligns with these efforts, as the spatial distribution of users can be leveraged to optimize training efficiency and resource allocation in real-world cellular networks.

The rest of the paper is organized as follows. Section~\ref{sec:sys_model} introduces the system model of federated learning and presents our realistic data partitioning approach. Section~\ref{sec:UE_selection} details the proposed clustering-based user selection strategy that leverages user metadata to mitigate data correlation. Simulation results and performance evaluations under various scenarios are presented in Section~\ref{sec:simulation}. Finally, Section~\ref{sec:conclusion} concludes the paper and discusses potential directions for future work.

\section{System Model}
\label{sec:sys_model}
\subsection{Federated learning}
\label{subsec:FL}
In this work, we consider a FL process where a BS or server tries to solve the following distributed optimization problem:
\begin{equation}
    \min~~F(w)= \sum_{k=1}^{N} p_k F_k(w), 
    \label{eq:global_loss}
\end{equation}
where $w$ are the model parameters. $N$ is the number of UEs. $p_k$ is the weight of UE$\#k$, where $p_k \geq 0$ and $\sum_{k=1}^Np_k=1$. $F_k(w)$ is the local cost function. Let $\mathcal{D}_k$ denotes the local dataset on UE$\#k$, and we have
\vspace{-1mm}
\begin{equation}
    F_k(w)=\frac{1}{D_k} \sum_{j=1}^{D_k} \ell(w;x_k^j,y_k^j),~~  k=1,2,\cdots,N, \label{eq:local_loss}
\end{equation}
where $D_k = |\mathcal{D}_k|$ is the number of samples in $\mathcal{D}_k$. $(x_k^j,y_k^j)$ is the $j$-th sample of UE$\#k$. $\ell(w;x_k^j,y_k^j)$ is the loss function on $(x_k^j,y_k^j)$.

We employ the \textit{FedAvg} algorithm \cite{mcmahan2017communication}. In the $t$-th FL training round, \textit{FedAvg} executes the following steps:
\subsubsection{\textbf{User selection and broadcasting}}
BS first selects a candidate set $S_t$ out of $K$ UEs, where $|S_t| = N$ and $N \leq K$. Each element of $S_t$ represents the index of the selected User. It then broadcasts the global model $w^{t-1}$ to users in $S_t$.
\subsubsection{\textbf{Local model updating}}
Each UE$\#k \in S_r$ updates the local model as follows:
\begin{equation}
    \begin{split}
    &\bar{w}_k^{0, t} = w^{t-1} \\
    &\bar{w}_k^{j, t} = \bar{w}_k^{j-1, t} - \eta\nabla F_k ( \bar{w}_k^{j-1, t}), \ j = 1,...,E, \\
    &w^r_t = \bar{w}_k^{E, t},
    \end{split}
    \label{eq:local_update}
\end{equation}
where $\eta > 0 $ is the learning rate. $E$ is the number of iterations. $w^{r}$ is the updated global model in $t$-th training round, and $w^{0}$ represents the initialized model at the beginning. $\bar{w}_k^{j, t}$ is the updated model parameters in the $j$-th iteration of $r$-th round at UE\#k. $\nabla F_k (\bar{w}_k^{j-1, t})$ is the gradient of UE\#$k$ at $\bar{w}_k^{j-1, t}$. $w^t_k$ is the model parameters to be uploaded in $t$-th round.
\subsubsection{\textbf{Aggregation}}
The selected UEs upload their local models $w^t_k$ to BS for aggregation:
\begin{equation}
    w^t = \frac{1}{\sum_{k \in S_t}n_k}\sum_{k \in S_t} n_k w_k^{t}.
\label{eq:global_agg}
\end{equation}

The user selection procedure will be further discussed in Section~\ref{sec:UE_selection}.

\subsection{Data partition}
\label{subsec:data_partition}
Consider a realistic scenario in sensor networks, such as fault detection and monitoring in a factory. Events occur in the environment and are sensed by all sensors within their coverage area. Each sensor collects all event samples in its sensing radius, resulting in overlapping data among neighboring sensors.

\begin{figure}
\centering
\includegraphics[width=0.95\linewidth]{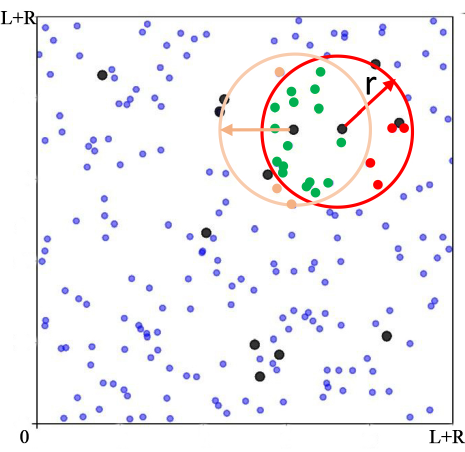}
\caption{Illustration of data partition model where points in color (blue, red, green, pink) are data points generated by HPPP over $\mathcal{S}_U = [-(L+R)/2,(L+R)/2]^2$, and the black points are UEs uniformly distributed in $\mathcal{S}_U = [-L/2, L/2]^2$.}
 \label{fig:data_partition}
\end{figure}

Inspired by \cite{kalor2018random,zheng2021stochastic,zheng2022stochastic}, we propose a new data partition model for FL simulation, illustrated in Fig.~\ref{fig:data_partition}. Let $\mathcal{S}_U = [-L/2,L/2]^2$ be a square region of size $L \times L$, representing the area where UEs are deployed. We assume that UEs are uniformly distributed over $\mathcal{S}_U$, i.e., $\mathbf{z}_i \sim \mathrm{Uniform}(\mathcal{S}_U)$ for UE~$i$ ($i=1,\dots,N$). 
Data points, representing events or samples in the environment, are generated according to a homogeneous Poisson point process (HPPP) with intensity $\lambda$ over a slightly larger square region $\mathcal{S}_D = [-(L+R)/2,(L+R)/2]^2$\footnote{The extension beyond $\mathcal{S}_U$ avoids edge effects, ensuring that UEs near the boundary still have a full sensing radius for collecting samples.}. Each data point corresponds to a single data sample. A UE collects a sample if the data point falls within its sensing radius $R$, i.e., the distance between the UE location $\mathbf{z}_i$ and the data point is no greater than $R$. 
Under this model, the number of samples collected by UE~$i$, denoted as $D_i$, follows a Poisson distribution:
\begin{equation}
    P\{D_i=n\} = \frac{(\lambda \pi R^2)^n}{n!} e^{-\lambda \pi R^2},
\end{equation}
with expected value
\begin{equation}
\label{eq:avg_num}
    \bar{D}_i = \lambda \pi R^2.
\end{equation}

This formulation explicitly captures two important aspects: (i) the heterogeneity of data quantity across UEs, and (ii) the potential overlap of data samples among neighboring UEs, which introduces natural data correlation.
Specifically, a single data sample may be collected by all UEs within its sensing radius $R$. Consequently, neighboring UEs that are closer together tend to exhibit higher data correlation due to larger overlap in their coverage areas, as illustrated by the green points in Fig.~\ref{fig:data_partition}. 
The dataset size can be adjusted by tuning the intensity of the data point distribution $\lambda$ and the sensing radius $R$. Similarly, the level of data correlation is influenced by the UE density ($N/L^2$) and $\lambda$. More realistic models could consider non-uniform UE distributions, heterogeneous sensing radii, or non-homogeneous point processes. For simplicity, this paper assumes uniform UE distribution, HPPP for data points, and a constant sensing radius $R$.

\section{User Selection Strategy Based on Metadata}
\label{sec:UE_selection} 
Data correlation can be understood as the ``overlap'' of data across users. High data correlation reduces data diversity, which has been shown to undermine FL performance if correlated UEs are selected \cite{sun2022user}. Since geographic distance is a key factor influencing data correlation between UEs, location should be considered in user selection for our scenario. 
Beyond location, many other factors contribute to data correlation in daily life. For example: (i) demographic attributes such as gender and age, e.g., young women may have similar preferences for cosmetics in online shopping; and (ii) temporal aspects, e.g., user activity patterns during events like the FIFA World Cup. These factors can be collectively described as ``\textit{metadata}'', which capture intrinsic characteristics that influence users' data patterns. As a result, users with similar metadata tend to exhibit high data correlation. Motivated by this observation, our goal is to select users in a way that minimizes data correlation, or equivalently, maximizes data diversity. For this reason, we propose our \textit{clustering-based user selection strategy}, which consists of two steps:

\subsubsection{\textbf{User Clustering}}
The central server or BS first collects the metadata of users, i.e. location information. It then partitions users into $N$ groups using a clustering algorithm, such as $k$-means. The clustering results are stored for user selection during the FL training phase.
\subsubsection{\textbf{User Selection}}
One user is randomly selected from each group to participate in the training process.

\begin{algorithm}
\caption{Federated Learning with Clustering-based User Selection}
\label{alg:FL_cluster}
    \begin{algorithmic}[1]
	\REQUIRE $\{\mathbf{M}_k\}$, $\{\mathcal{D}_k\}$, $T$ and $N$.
	\ENSURE $w^*$.
	\STATE Initialize the global model $w^0$
	\STATE $\{\mathbf{I}_n\}$  =  \textbf{User\_Clustering$\left(\{ \mathbf{M}_k \}\right)$}
	\FOR{each round $t = 1, \ldots, T$}
	   \STATE $\mathbf{S}_t$ = \textbf{User\_Selection$\left(\{\mathbf{I}_n\}\right)$}
	   \STATE $w^t$ = \textbf{Model\_Training$(\{\mathcal{D}_k\}, \mathbf{S}_t)$}
	\ENDFOR

        \STATE $w^*=w^T$\\[4pt]

	\STATE \textbf{User\_Clustering:}
        \STATE $\{\mathbf{I}_n\}$ = K-means$\left(\{\mathbf{M}_k\}, N\right)$
        \RETURN $\{\mathbf{I}_n\}$\\[4pt]
        
        \STATE \textbf{User\_Selection$\left(\{\mathbf{I}_n\}\right)$}:
        \STATE $\mathbf{S}_t=\{\}$
        \FOR{each group $n = 0, 1, \ldots, N$}
          \STATE Randomly choose $s_n$ where $s_n \in \mathbf{I}_n$
          \STATE $\mathbf{S}_t = \{\mathbf{S}_t, s_n\}$
        \ENDFOR
        \RETURN $\mathbf{S}_t$\\[4pt]
    
        \STATE $w^t$ = \textbf{Model\_Training$(\{\mathcal{D}_k\}, \mathbf{S}_t)$}:
        \STATE The server and users follows the step 1)--3) in Section~\ref{subsec:FL}
        \RETURN $w^t$\\[4pt]
	\end{algorithmic}
$\{ \mathbf{M}_k \} = \{\mathbf{M}_1, \mathbf{M}_2, \dots, \mathbf{M}_K\}$ where $\mathbf{M}_k$ denotes the metadata of $k$-th user;\\
$T$ is the maximum number of training round;\\
$N$ is the number of user groups;\\
$\{\mathbf{I}_n\} = \{\mathbf{I}_1, \mathbf{I}_2, \ldots, \mathbf{I}_N\}$ where $\mathbf{I}_n$ denotes the indices of users within $n$-th group.

\label{alg:fed_cluster}
\end{algorithm}

The overall procedure is summarized in Algorithm~\ref{alg:FL_cluster}, where \textbf{User\_Clustering} is executed only once, while \textbf{User\_Selection} and \textbf{Model\_Training} are repeated until the global model converges:
As a result of the \textbf{User\_Clustering} step, users are partitioned into $N$ groups. The value of $N$ can be chosen based on a rule of thumb or practical considerations, e.g., equal to the number of available radio resources. Since each user uploads metadata only once, the communication overhead is negligible. Moreover, some metadata such as UE location are natively available to 3GPP network functions (e.g., via Location Services and the Network Exposure Function), which enables metadata-aware client selection in cellular deployments.
In the \textbf{User\_Selection} step, we rely on the intuition that users within the same group possess similar data samples, whereas users from different groups tend to have distinct or independent samples. Accordingly, one user is selected from each group to participate in FL training.

\section{Simulation Results}
\label{sec:simulation}
In this section, we conduct experiments on two real-world datasets, namely Fashion-MNIST (FMNIST)~\cite{xiao2017fashion} and CIFAR-10~\cite{krizhevsky2009learning}, to evaluate the performance of our proposed framework and investigate the impact of key hyperparameters.  

\subsection{Experiment Setup}  
\textbf{Dataset:} As described in Section~\ref{subsec:data_partition}, we assume the coverage area of a BS is a $10 \times 10$ square, where $K$ users are uniformly distributed and data points follow a homogeneous Poisson point process (HPPP) with intensity $\lambda$. The sensing radius is set to $R=2$. In addition, we consider two scenarios as in~\cite{mcmahan2017communication}:

\subsubsection{\textbf{IID}} 
We assume that data points are randomly shuffled and distributed among users, resulting in no label skewness across the dataset.

\subsubsection{\textbf{Non-IID}} 
\begin{figure}[h]
	\centering
	\includegraphics[width=0.95\linewidth]{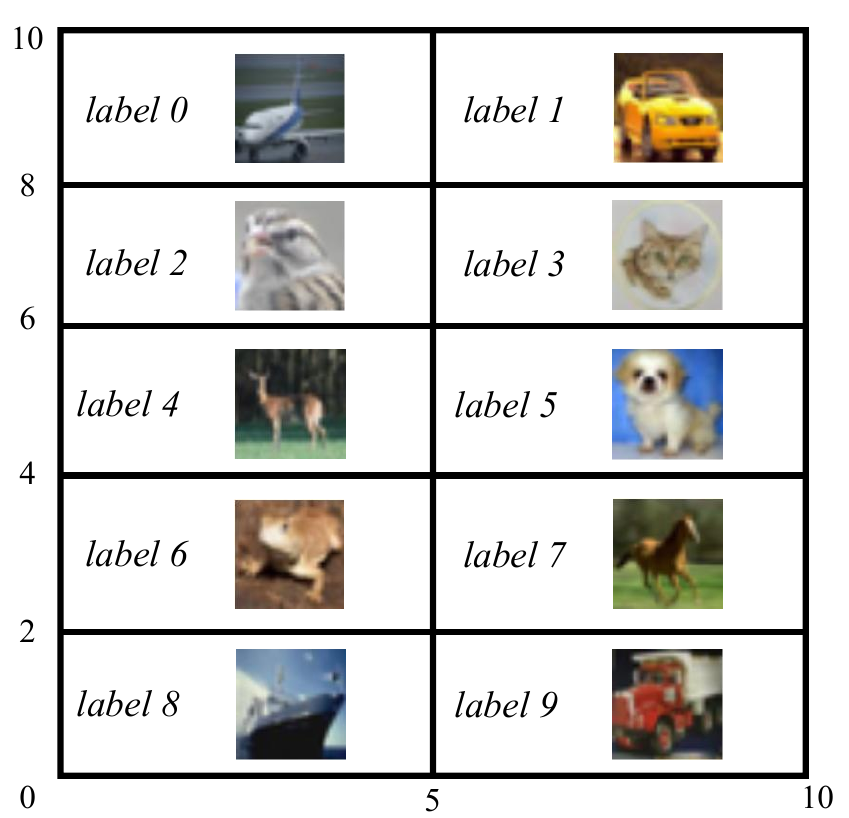}
	\caption{Non-IID data partition.}
	\label{fig: non_iid_setting}
\end{figure}

In this scenario, the label of each data point is determined by its spatial location, as illustrated in Fig.~\ref{fig: non_iid_setting}. The square coverage area is divided into $10$ equal rectangular regions, with each region assigned a unique label. Consequently, data points within the same region share the same label, leading to extreme label skewness among users located in different areas.

% and we adjust  that is the parameters of PPP method mentioned in Section III to simulate the real WFL scenario, i.e., simulating the distribution of device and data, and the data partition method is also mentioned in Section III. We also assume that the label of data has strong correlation with the location as shown in Fig.\ref{fig: non_iid_setting}, I

% t means the distribution of data that is collected by devices in different location is different and this phenomenon causes the data skewness.

\textbf{FL Training}: 
For each user, a Multilayer Perceptron (MLP) is employed as the local model for the image classification task. The MLP architecture is $784 \times 200 + 200 \times 10$ for FMNIST and $3072 \times 200 + 200 \times 10$ for CIFAR-10. Stochastic Gradient Descent (SGD) is used for optimization with a learning rate of $\eta=0.001$. We set the number of local epochs to $E=1$ and the batch size to $b=32$. Test accuracy and training loss are used as evaluation metrics. For baseline comparison, we adopt a random sampling strategy for user selection.

\subsection{Simulation results}
In this subsection, we evaluate the performance of our proposed method under various experimental settings.

\subsubsection{\textbf{Performance evaluation of FMNIST and CIFAR-10 in IID scenario}} 
We set the intensity of data points as $\lambda = 500$, the total number of users $K=10,000$, the number of clusters $N_c=5$, and the number of communication rounds $T=500$. We further assume that the number of selected users in each round equals the number of groups, i.e., $N_u=N_c=5$. 

\begin{figure}[h]
	\centering
	\subfloat[\scriptsize{CIFAR-10} Test Accuracy] {\includegraphics[width=0.465\linewidth]{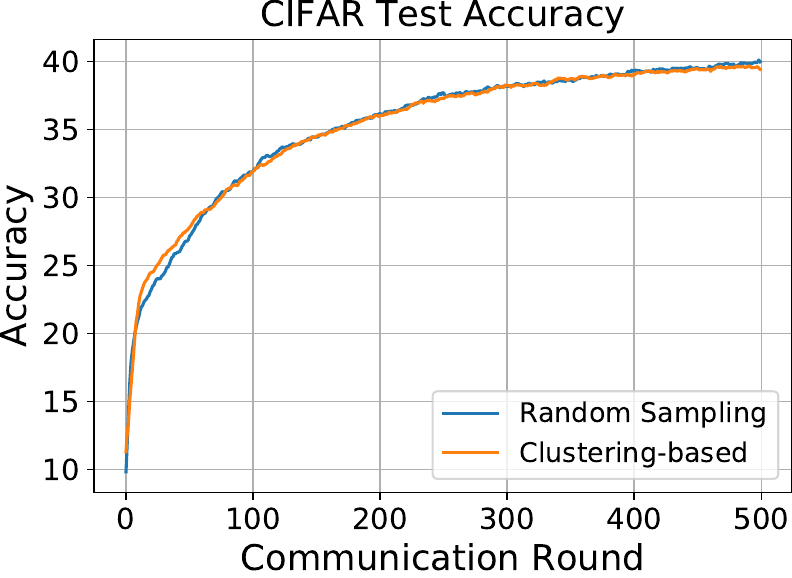} \label{3-1}}
	\subfloat[\scriptsize{CIFAR-10} Training Loss]{\includegraphics[width=0.482\linewidth]{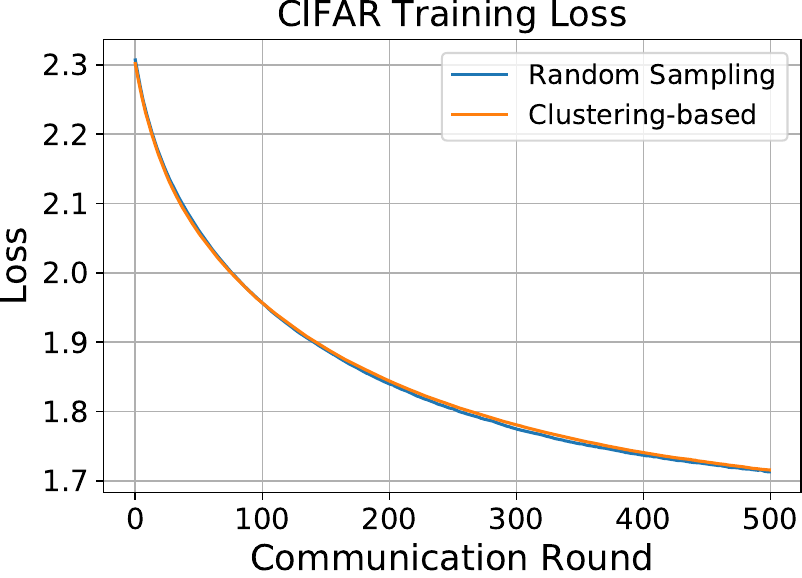} \label{3-2}}
  \hfill
	\subfloat[\scriptsize{FMNIST} Test Accuracy]{\includegraphics[width=0.475\linewidth]{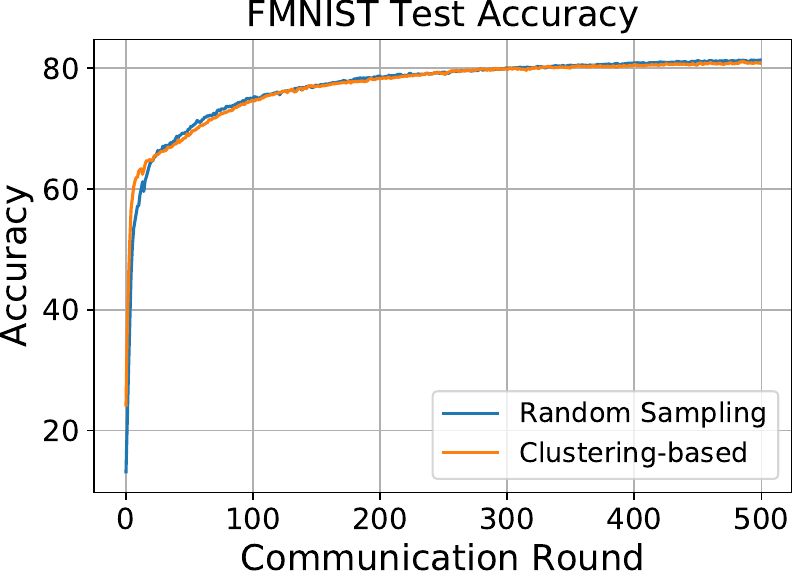}
		\label{3-3}}
	\subfloat[\scriptsize{FMNIST} Training Loss]{\includegraphics[width=0.48\linewidth]{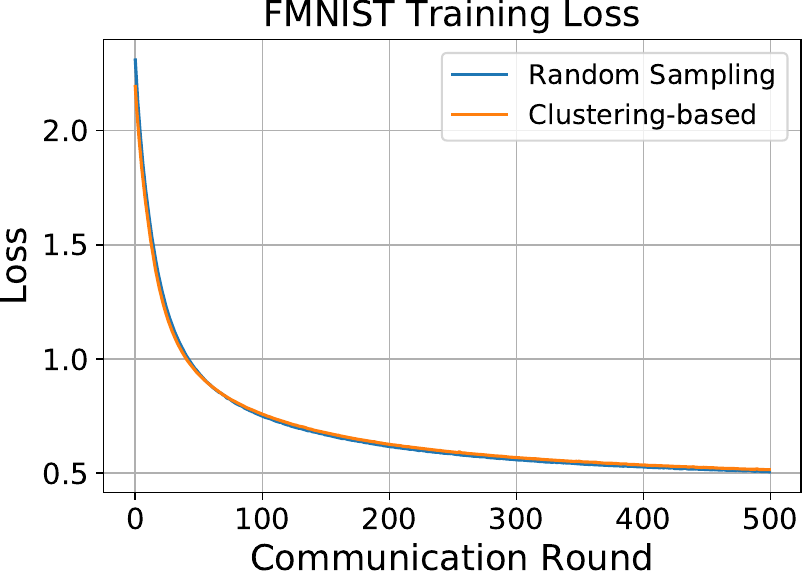}
		\label{3-4}}
	\caption{Performance evaluation of CIFAR-10 and FMNIST in IID scenario}
	\label{fig: iid_experiment}
\end{figure}

The results are shown in Fig.~\ref{fig: iid_experiment}. 
As observed, there is little difference between our proposed clustering-based user selection method and random sampling. This is because each user has a sufficiently large number of samples (average $6280$ according to \eqref{eq:avg_num}) covering all categories in the IID setting. Therefore, the performance gain from user selection is negligible in this case.

\subsubsection{\textbf{Performance evaluation of FMNIST and CIFAR-10 in non-IID scenario}}
We keep the parameters the same as in the IID scenario, i.e., $\lambda = 500$, $K=10,000$, $N_u=N_c=5$, and $T=500$. 

\begin{figure}[h]
	\centering
	\subfloat[\scriptsize{CIFAR-10} Test Accuracy] {\includegraphics[width=0.48\linewidth]{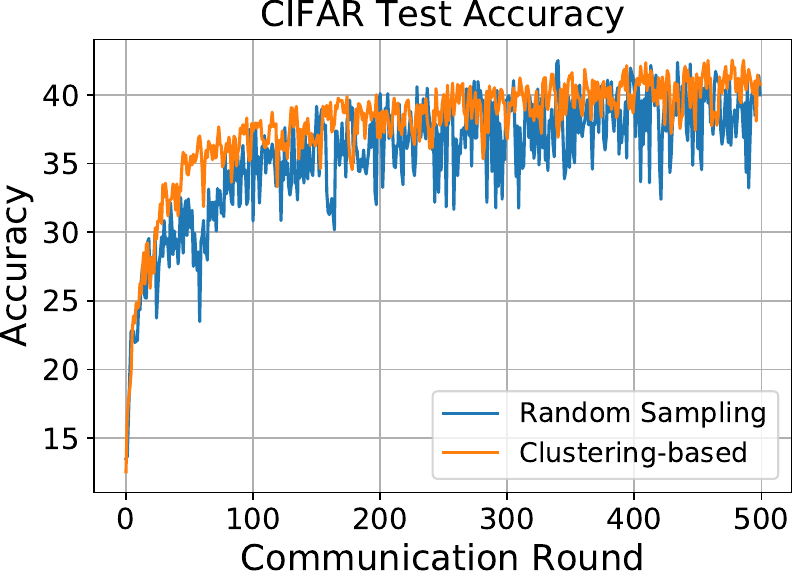} \label{4-1}}
	\subfloat[\scriptsize{CIFAR-10} Training Loss]{\includegraphics[width=0.482\linewidth]{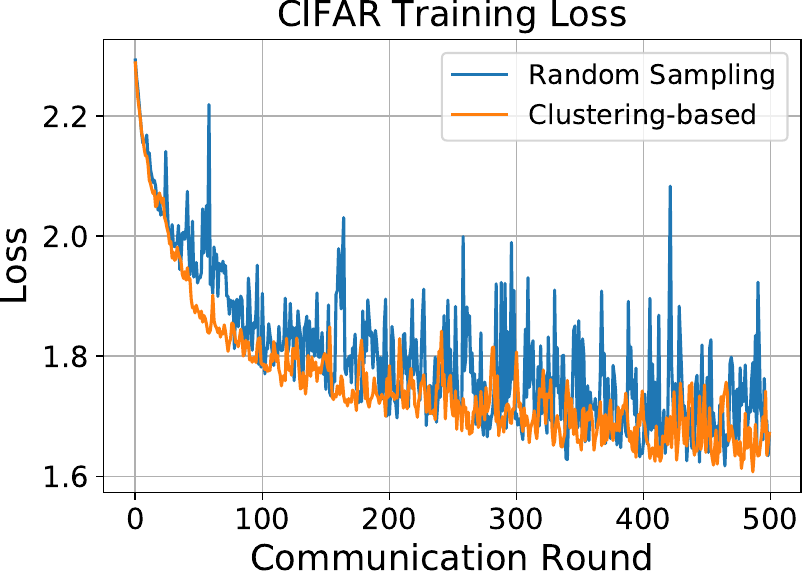} \label{4-2}}
  \hfill
	\subfloat[\scriptsize{FMNIST} Test Accuracy]{\includegraphics[width=0.475\linewidth]{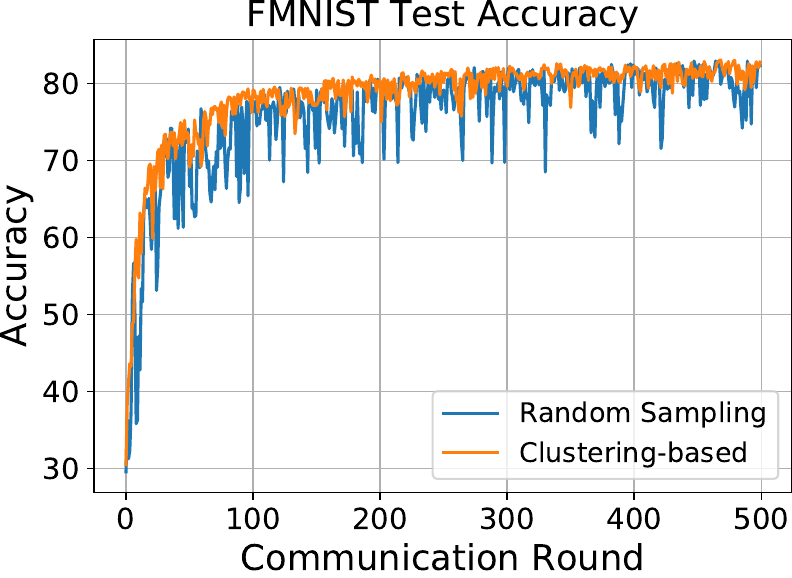}
		\label{4-3}}
	\subfloat[\scriptsize{FMNIST} Training Loss]{\includegraphics[width=0.48\linewidth]{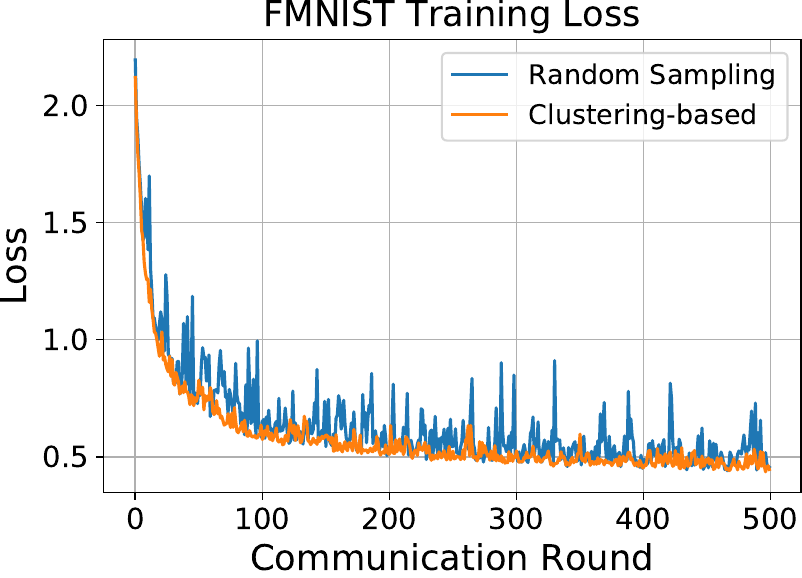}
		\label{4-4}}
	\caption{Performance evaluation of CIFAR-10 and FMNIST in non-IID scenario}
	\label{fig: non_iid_experiment}
\end{figure}

As shown in Fig.~\ref{fig: non_iid_experiment}, our clustering-based user selection method outperforms random sampling in terms of test accuracy, training loss, and stability. In non-IID scenarios, random sampling has a higher probability of selecting neighboring users with similar labels, leading to label skewness in each training round and causing severe fluctuations during training. Moreover, while the improvement may appear subtle in the figure, our method consistently achieves approximately $1\%$ higher test accuracy compared to random sampling. This improvement can be attributed to a more balanced label distribution across selected users in each round, leading to more stable and efficient SGD convergence.

\subsubsection{\textbf{FL performance under different total user numbers}} 
We set $\lambda = 500$, $N_u=N_c=5$ and $T=200$, and make comparison of the performance under $K=20$ and $K=10,000$.

\begin{figure}[h]
	\centering
	\large
	\subfloat[\scriptsize{CIFAR, $K=20$}] {\includegraphics[width=0.45\linewidth]{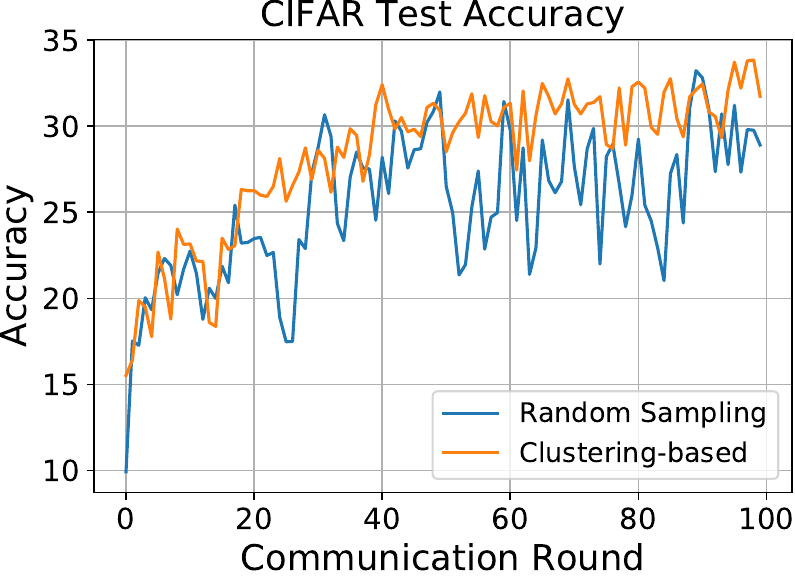}
		\label{Fig:Nc_20}}
	\subfloat[\scriptsize{CIFAR, $K=10000$}] 
	{\includegraphics[width=0.45\linewidth]{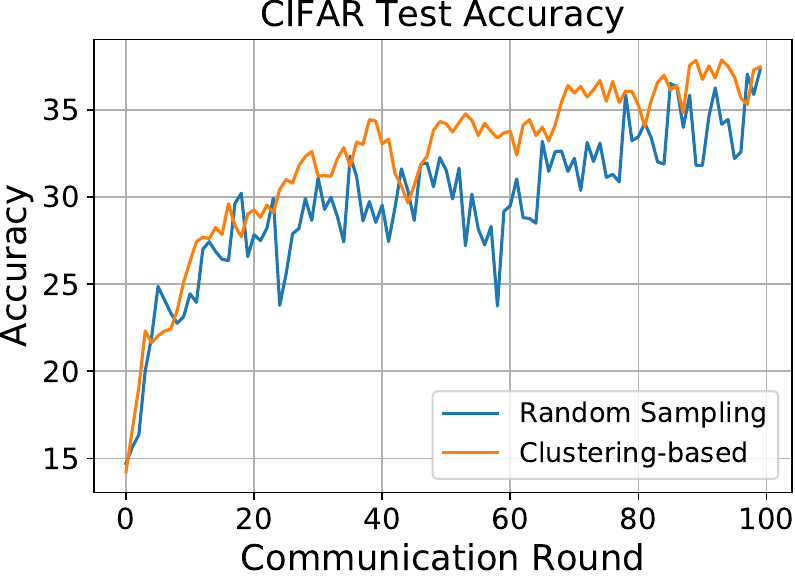}
    \label{Fig:Nc_10000}}
    \hfill
        \subfloat[\scriptsize{FMNIST, $K=20$}] 
	{\includegraphics[width=0.45\linewidth]{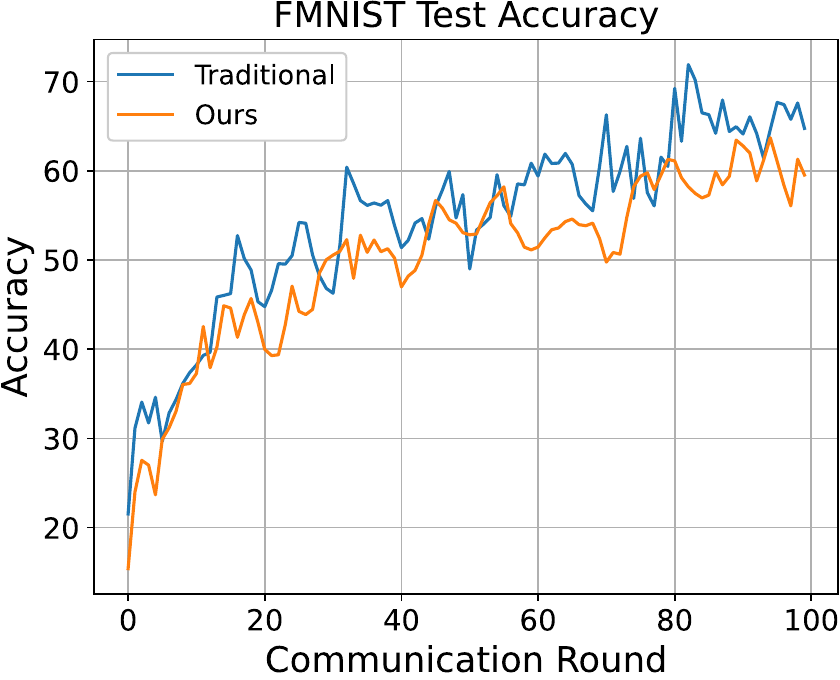}
    \label{Fig:Nc_10000}}
        \subfloat[\scriptsize{FMNIST, $K=10000$}] 
	{\includegraphics[width=0.45\linewidth]{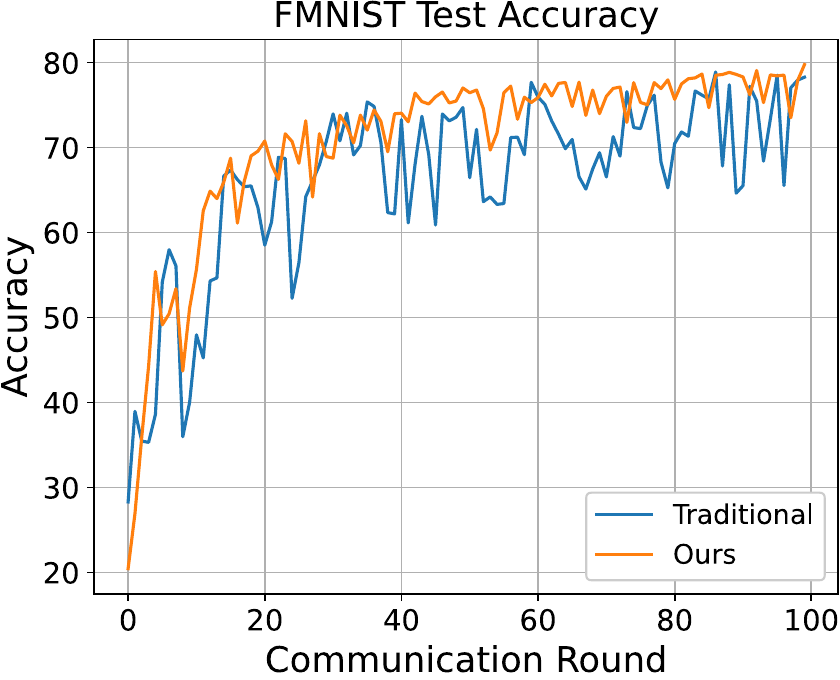}
    \label{Fig:Nc_10000}}
	\caption{Impact of total user number $K$ on FL performance.}
	\label{fig: different_users_number}
\end{figure}

As shown in Fig.~\ref{fig: different_users_number}, the performance of our proposed framework improves as $K$ increases. This is because a larger number of users leads to a denser and more uniform spatial distribution of UEs, which allows our clustering-based user selection to more effectively choose spatially separated users with lower data correlation. Consequently, the model benefits from more diverse data in each training round, improving both convergence stability and overall accuracy.

\subsubsection{\textbf{{FL performance under different selected user number(or cluster number)}}} 
\begin{figure}[h]
	\centering
	\subfloat[\scriptsize{Test Accuracy}, $N_c = 100$] {\includegraphics[width=0.465\linewidth]{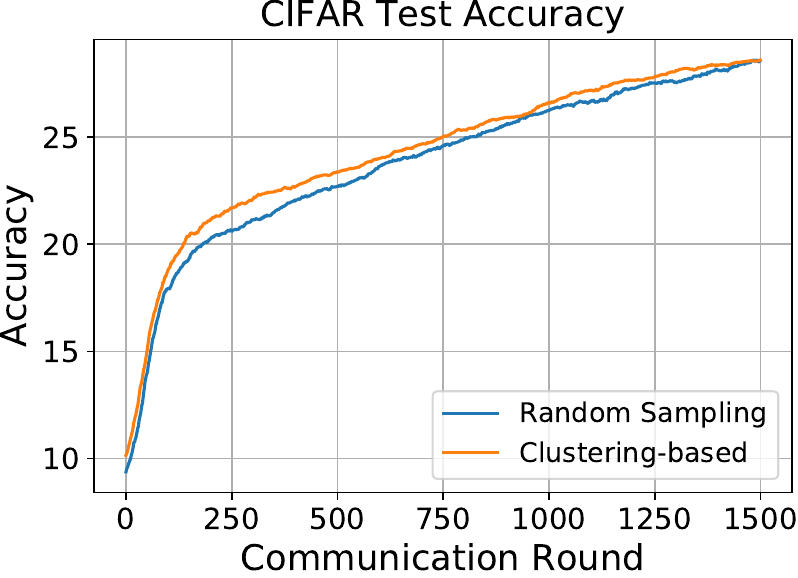} \label{6-1}}
	\subfloat[\scriptsize{Training Loss}, $N_c = 100$]{\includegraphics[width=0.482\linewidth]{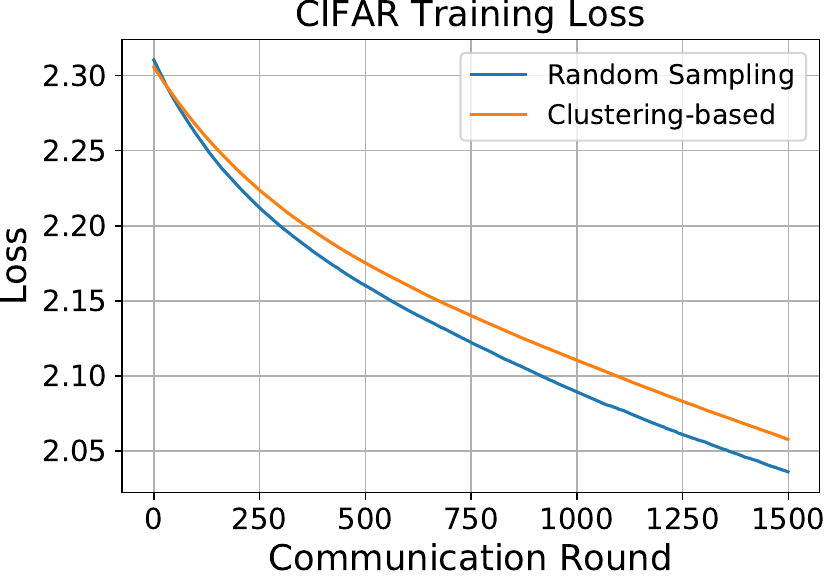} \label{6-2}}
  \hfill
	\subfloat[\scriptsize{Test Accuracy}, $N_c = 20$]{\includegraphics[width=0.465\linewidth]{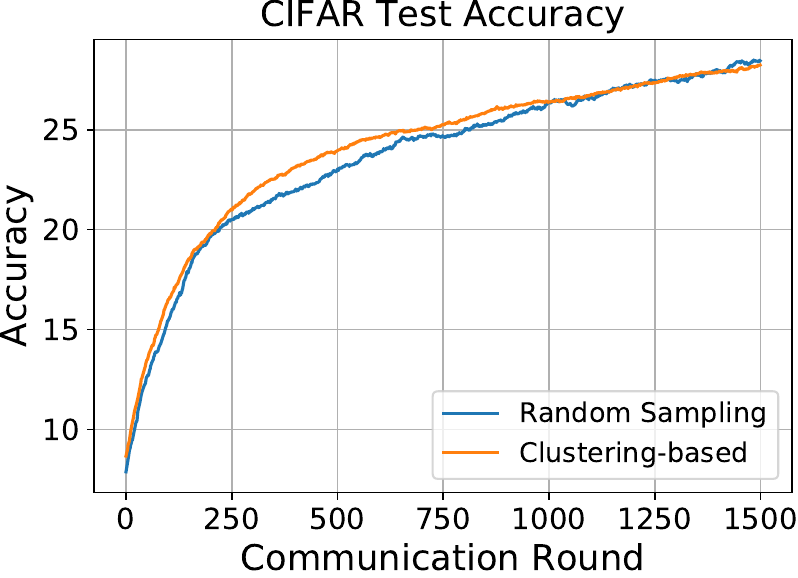}
		\label{6-3}}
	\subfloat[\scriptsize{Training Loss}, $N_c = 20$]{\includegraphics[width=0.482\linewidth]{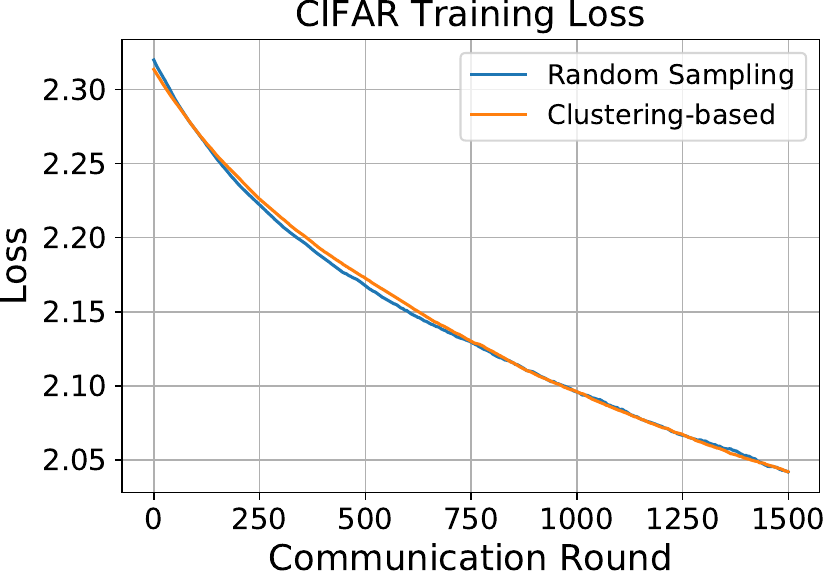}
		\label{6-4}}
  \hfill
	\subfloat[\scriptsize{Test Accuracy}, $N_c = 10$]{\includegraphics[width=0.465\linewidth]{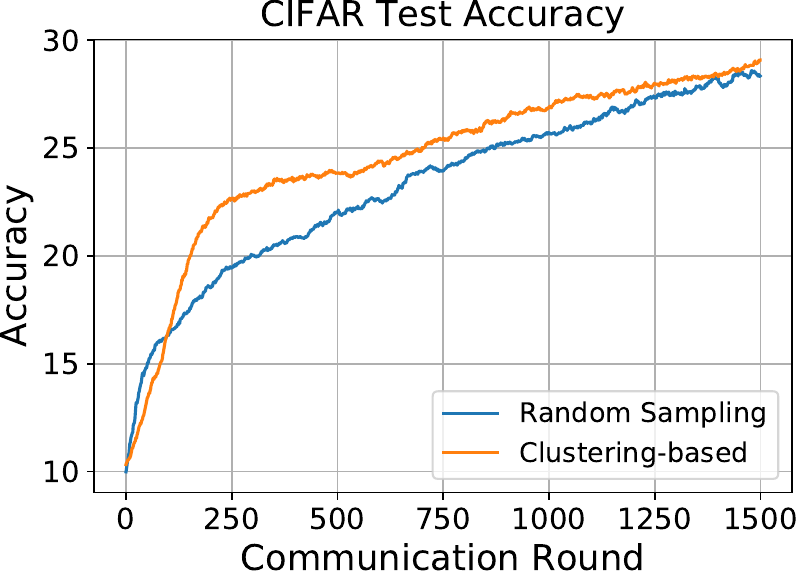}
		\label{6-5}}
	\subfloat[\scriptsize{Training Loss}, $N_c = 10$]{\includegraphics[width=0.482\linewidth]{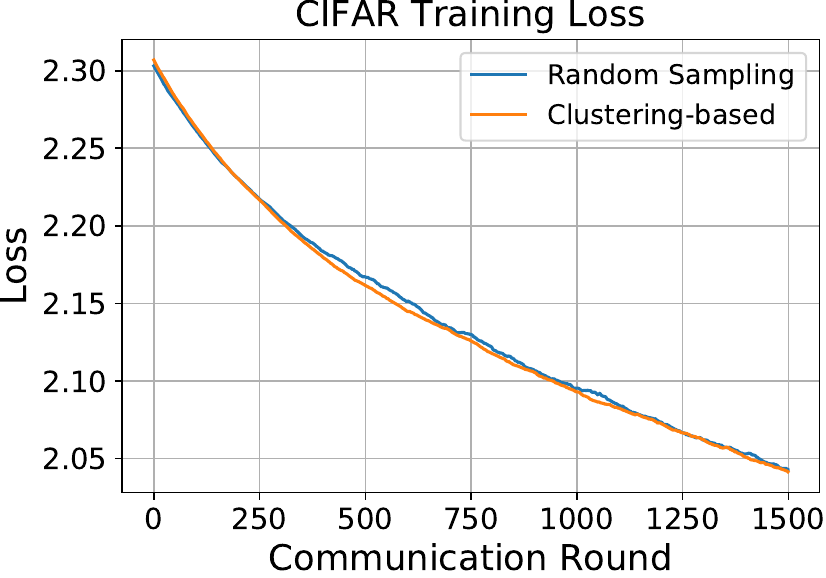}
		\label{6-6}}
	\caption{Impact of different selected numbers $N_c$ on FL performance.}
	\label{fig: different_selected_number}
\end{figure}

 We set the parameters as follows: $\lambda=500$, $K=200$, and $T$ is 1500. As shown in Fig.~\ref{fig: different_selected_number}, when $N_c$ equals $100$ or $20$, the performance of our proposed method is similar to that of the traditional method. However, when $N_c$ decreases to $10$, our method shows a significant performance improvement. 
This can be explained as follows: when more users are selected in each round, the training data naturally contain more labels, resulting in similar data distributions between the traditional and proposed methods. Consequently, their performance differences become negligible. 
In contrast, when the number of selected users is small, the traditional method suffers from label deficiency, whereas our clustering-based selection maintains label diversity and balanced data distribution. Therefore, our proposed method achieves better performance when the number of selected users (or clusters) is small.

\subsection{Analysis}
Extensive experiments on FMNIST and CIFAR-10 demonstrated the effectiveness of the proposed framework. Specifically, the results show that: 

1) \textbf{Superior performance in non-IID scenarios:} Compared with random user selection, our method effectively addresses the lack of label information and data overlapping, achieving higher global model accuracy during training.  

2) \textbf{Improved stability and convergence:} By balancing the label distribution among selected users, the proposed framework reduces fluctuations in training and ensures smoother convergence, particularly in non-IID settings.  

3) \textbf{Effectiveness under limited user participation:} The method shows significant benefits when few users are selected per round, preserving label diversity and improving convergence, while performing similarly to traditional methods under IID settings.

4) \textbf{Limitations:} The framework’s effectiveness is sensitive to the number of clusters and the degree of data heterogeneity; when data distribution is already close to IID, the benefit of clustering-based selection becomes negligible.

\section{Conclusion and Discussion}
\label{sec:conclusion}
This paper proposed a metadata-driven federated learning framework that addresses the challenges of unrealistic data partitioning and user data correlation in conventional FL simulations. By introducing a data partition model based on the homogeneous Poisson point process (HPPP) and a clustering-based user selection strategy leveraging metadata, the framework improves model accuracy, stability, and convergence, especially in non-IID scenarios or under limited user participation.

Beyond simulation results, the proposed framework has practical implications for large-scale cellular and edge networks:

1) \textbf{Alignment with 3GPP standards:} User metadata, particularly location information, is natively supported by 3GPP network functions such as Location Services and the Network Exposure Function. This makes metadata-aware client selection feasible in real-world deployments and provides a concrete approach for supporting FL applications in cellular networks.

2) \textbf{Scalability and industrial relevance:} The method reduces training fluctuations and communication overhead, enabling efficient FL in large-scale IoT and mobile edge computing scenarios. By preserving label diversity and mitigating data correlation, the framework supports high-quality distributed learning even when only a subset of users participates in each round.

3) \textbf{Future directions:} Adaptive clustering strategies, incorporation of richer metadata types (e.g., temporal activity patterns, demographic features), and integration with large-scale wireless and edge networks will be explored to further enhance FL performance and industrial deployment readiness.

Overall, this work bridges the gap between realistic FL simulation and practical network deployment, providing both theoretical insights and actionable guidelines for standardization and industrial applications.

\section*{Acknowledgment}
This work is fully supported by Sony Research and Development Center and Sony (China) Ltd. The authors would like to thank Zhang Ke from Waseda University and Ma Shiyao from Southwest University for their contributions as interns at Sony.

\bibliographystyle{IEEEtran}
\bibliography{ref}

\vfill

\end{document}